\documentclass[12pt]{article}
\usepackage{amsfonts,amsmath,amssymb,mathrsfs,hyperref,url,tikz}

\title{Bohmian Mechanics at Space-Time Singularities. I.\\
   Timelike Singularities}

\author{
Roderich Tumulka\footnote{Mathematisches Institut, Eberhard-Karls-Universit\"at, Auf der Morgenstelle 10, 72076 T\"ubingen, Germany. E-mail: {\tt roderich.tumulka@uni-tuebingen.de}}
}
\date{July 25, 2019}

\newcommand{\Hilbert}{\mathscr{H}}
\newcommand{\conf}{\mathcal{Q}}

\renewcommand{\Im}{\mathrm{Im}}
\newcommand{\PPP}{\mathbb{P}}
\newcommand{\RRR}{\mathbb{R}}
\newcommand{\CCC}{\mathbb{C}}
\newcommand{\SSS}{\mathbb{S}}

\newcommand{\vx}{\boldsymbol{x}}

\newcommand{\st}{\mathscr{M}}
\newcommand{\boun}{\partial \st}
\newcommand{\inte}{\st^\circ}
\newcommand{\boundary}{\partial\conf}
\newcommand{\interior}{\conf^\circ}
\newcommand{\D}{d}
\newcommand{\dest}{f}
\newcommand{\foliation}{\mathscr{F}}
\newcommand{\sing}{\mathscr{S}} % singularity
\newcommand{\spin}{S} % spin space
\newcommand{\vol}{\mu}
\newcommand{\area}{\lambda}
\newcommand{\antidest}{\nu}

\newcommand{\be}{\begin{equation}}
\newcommand{\ee}{\end{equation}}

\begin{document}
\maketitle
\begin{abstract}
We develop an extension of Bohmian mechanics to a curved background space-time containing a singularity. The present paper focuses on timelike singularities. We use the naked timelike singularity of the super-critical Reissner--Nordstr\"om geometry as an example. 
While one could impose boundary conditions at the singularity that would
prevent the particles from falling into the singularity, we are interested here in the case
in which particles have positive probability to hit the singularity and get annihilated.
The wish for reversibility, equivariance, and the Markov property then dictates
that particles must also be created by the singularity, and indeed dictates
the rate at which this must occur. That is, a stochastic law prescribes what
comes out of the singularity. 
We specify explicit equations of a model involving an interior--boundary condition on the
wave function at the singularity, which can be used also in other versions of quantum 
theory besides Bohmian mechanics. As the resulting theory involves particle creation and annihilation, it can be regarded as a quantum field theory, and the stochastic process for the Bohmian particles is analogous to Bell-type quantum field theories.

\medskip

  \noindent 
  Key words: 
  interior-boundary condition;
  quantum theory in curved background space-time; 
  Reissner--Nordstr\"om space-time geometry; 
  sto\-chastic jump process;
  particle creation and annihilation;
  Bohmian trajectories.
\end{abstract}

\section{Introduction}

We study the construction and behavior of quantum mechanics in a space-time with timelike singularities using Bohmian mechanics (also known as pilot-wave theory), a version of quantum mechanics with particle world lines. We develop an explicit (mathematically non-rigorous) model, defining trajectories on a given background space-time geometry, using the Dirac equation, Bohm's law of motion, and a novel interior--boundary condition at the singularity that governs the creation and annihilation of particles at the singularity. As a case study, we consider the Reissner--Nordstr\"om geometry in the ``charge $>$ mass'' regime (the ``super-critical'' case), which features a naked timelike singularity. 
We use this geometry as a background in which the Bohmian particles move and get created and annihilated, while we neglect any change in the metric that these particles might be expected to cause. 
The Bohmian approach naturally leads us, in a space-time with a timelike singularity, to a quantum field theory in which particles get created and annihilated at the singularity. 

Works on quantum field theory in curved space-time often focus on globally hyperbolic space-times (e.g., \cite{FR16}), which are free of timelike singularities. 
Prior works considering a quantum evolution on a curved background space-time with a timelike singularity include \cite{DD78,Wal80,HM95,Bel98,HI99,IW,ST04,Fra16}, but they did not consider particle creation at the singularity.

Timelike singularities are usually regarded as unreasonable because they seem to make physics lawless: Given that the world line of a particle reaching the singularity $\sing$ ends there, any matter could disappear at $\sing$. This would seem to violate the conservation of mass-energy; moreover, and much worse, by time reversal symmetry, any matter could come out of $\sing$, so the theory should become not merely indeterministic but lawless. Now our model shows that the situation need not be that bad. The model is stochastic but not lawless: a quite simple boundary condition on the wave function and a natural prescription governing the stochastic creation of Bohmian particles fill exactly the gaps in the laws caused by the singularity while preserving time reversal symmetry. Also, the evolution of the wave function is governed by a Hamiltonian, so energy is arguably conserved.

We do not contribute here to a different question about singularities that is often asked: whether cosmic censorship holds, i.e., whether it is true that, generically, naked timelike singularities do not form. Some works \cite{DD78,Bel98} have contributed in this direction by arguing that the super-critical Reissner--Nordstr\"om metric is unstable, i.e., that various mechanisms would quickly make it decay into a sub-critical Reissner--Nordstr\"om metric (in which the timelike singularity is hidden behind a horizon), and in fact that the super-critical one would probably not form in the first place (for example through gravitational collapse). 

We focus here on a different observation: that physics may well make sense on a space-time with timelike singularity. Results in this direction have been obtained before by \cite{Wal80,HM95,HI99,IW,ST04,Fra16}, but they focused on certain kinds of boundary conditions on the wave function that would avoid any current into or out of the singularity $\sing$. Here, in contrast, we focus on the natural possibility of particle creation and annihilation at $\sing$.

\subsection{Bohmian Mechanics}

Bohmian mechanics \cite{Bohm52,Gol01} possesses a natural extension to relativistic space-time if a preferred foliation of space-time into spacelike hypersurfaces (called the \emph{time foliation}) is granted \cite{HBD}. This extension has also been formulated for curved space-time geometries \cite{3forms,Tum06d} and for ones with spacelike singularities \cite{Tum08b}, but not yet for geometries with timelike singularities. While horizons present no difficulty, singularities require further work to define the theory: Basically, we have to specify what happens when a particle hits the singularity, since at this point the law of motion is no longer defined. The possibility we consider here is that the particle gets annihilated: that is, if the system consisted of $n$ particles, then the world line of the particle that has arrived at the singularity ends there, while the other $n-1$ particles, which are not at the singularity and thus have no reason to vanish, continue to move according to Bohm's law of motion. To make this possible, we need wave functions from Fock space, as always when particles can get created or annihilated. Further considerations then naturally lead us to specific equations, defining a Bohm-type theory.

\subsection{Particle Creation}

In our model, particles can not only be annihilated but also be \emph{created} at the singularity. Indeed, this occurs at random times distributed in a way determined by a law involving the wave function. As a consequence, the theory is no longer deterministic; instead, the evolution of the particles can be described by a Markovian stochastic process in the appropriate configuration space. This process has similarities with processes used earlier in ``Bell-type quantum field theories'' \cite{crea2B}, a stochastic version of Bohmian mechanics for quantum field theories that we describe briefly in Section~\ref{sec:BTQFT}. The process is defined by two equations: Bohm's equation of motion and an equation that determines the jump rate for creation of particles at the singularity. The latter is uniquely fixed if we strive for the following properties: reversibility, equivariance, and the Markov property. In classical mechanics, in contrast, there is no good reason for postulating any law about when and where particle world lines begin at the singularity. Thus, both Bohmian and classical mechanics become \emph{indeterministic} in the presence of timelike singularities, but in different ways: while classical mechanics becomes \emph{lawless}, Bohmian mechanics becomes \emph{stochastic}. A timelike singularity causes concern in classical mechanics, but not in quantum mechanics.

It is an interesting moral from our model that particle creation and annihilation, one of the characteristic features of quantum field theory, can arise from the presence of timelike singularities. This invites speculation about whether quantum particles capable of emitting other particles might actually contain timelike singularities. After all, the classical gravitational field of a point particle with the charge and the mass of an electron is exactly the Reissner--Nordstr\"om geometry we are considering. This does not prove that the ultimate theory of the electron, including quantum gravity, will attribute a timelike singularity to it, but at least that seems like an intriguing possibility.

The concrete model we develop is based on the Dirac equation and the Reissner--Nordstr\"om geometry, but we expect that similar models can be constructed on the basis of other wave equations 
and other space-times containing timelike singularities. We have discussed Bohmian mechanics at spacelike singularities elsewhere \cite{Tum08b}; lightlike singularities behave, for our purposes, like spacelike ones. (A general singularity may well, like a general 3-surface, have different regions in which it is timelike, spacelike, and lightlike. Correspondingly, our separate studies of timelike and spacelike singularities apply separately to the timelike and the spacelike subset of the singularity.)

The Dirac equation is known to have solutions of negative energy that are considered unphysical; indeed, the Hamiltonian is unbounded from below. We will not worry about this here, but rather treat the negative energy states as if they were physical.

\subsection{Motivation}

The investigations in this paper have several motivations and lead to benefits in several areas: 
\begin{itemize}
\item It is a natural part of the research program on Bohmian mechanics to extend the theory to more general quantum theories, to all kinds of settings. To the extent that we have reason to believe that singularities exist in our universe, we obtain here a more appropriate version of Bohmian mechanics. Concerning techniques of constructing Bohm-type models, we find that the Bohm-type law of motion proposed by D\"urr \emph{et al.}~\cite{HBD} for relativistic space-time with a foliation works unproblematically even under the extreme conditions at a singularity; that not every spacelike foliation is equally good for that purpose (as we discuss in Section~\ref{sec:foliation}); that the stochastic approach to particle creation developed in ``Bell-type quantum field theories'' \cite{crea2B} arises naturally in this context; and that particle annihilation may well be deterministic and particle creation stochastic, even though the evolution is reversible.

\item Since Bohmian mechanics is a precise and unambiguous version of quantum mechanics, it may serve as a tool for studying quantum mechanics in curved space-time. Thus, our study can as well be regarded as one on \emph{quantum mechanics at space-time singularities}. Concretely, we propose a novel unitary evolution of the wave function in the presence of a timelike singularity, based on the Dirac equation and an interior--boundary condition at the singularity (independently of the Bohmian viewpoint). Interior--boundary conditions relate values of the wave function at the boundary of the $n$-particle sector of configuration space to the value at an interior point in (usually) the $(n-1)$-particle sector. Such conditions have been suggested also in flat space-time \cite{Mosh51a,Tho84,Yaf92,GT05b,TT15a,TT15b,KS16,co1}, sometimes for the purpose of devising Hamiltonians with particle creation and annihilation at a point-shaped source; interior--boundary conditions for the Dirac equation have been considered in \cite{STT18,LN18}. While interior--boundary conditions for the non-relativistic Schr\"odinger equation sometimes have the advantage of being free of ultraviolet divergence \cite{LSTT}, this is not expected in our model; we leave aside the question of ultraviolet divergence.

\item Several interesting features arise in the theory we develop: Timelike singularities emit particles; we can specify the law for how they do it; certain traits of quantum field theory (such as Fock space) arise from the presence of singularities; timelike singularities need not be unphysical. 
\end{itemize}

I emphasize that the theory developed in this paper does not merely add Bohmian trajectories to otherwise known ``orthodox'' quantum theories. Instead, the Bohmian viewpoint helps define what the evolution equation should be, also for the wave function. Indeed, the Fock space evolution corresponding to particle creation at a timelike singularity \emph{could have been, but has not been}, considered in an orthodox framework. The reason why it has not been considered before is, I think, twofold: First, since singularities often lie behind event horizons, and since orthodox quantum physics tends not to ask what actually happens but rather what observers see, it is common to trace out the degrees of freedom behind event horizons, which means to ignore the phenomena this paper is concerned with. Second, in orthodox quantum physics it is common to focus on the scattering regime, i.e., on the asymptotic distribution of particles a long time after their interaction. For defining a Bohm-type theory, in contrast, it is essential to define the trajectories, which may well depend on the particles hidden behind the event horizon, and this forces us to pay attention to what the time evolution laws are. Of course, also independently of the Bohmian approach one may feel that real events can take place inside black holes and should therefore not be excluded from the description in a physical theory; as Dafermos \cite{Daf15} once put it, ``People in black holes have rights, too!'' Note further that a scattering theory can only be formulated in an asymptotically flat space-time, while our approach does not require asymptotic flatness or stationarity, even though the Reissner--Nordstr\"om geometry happens to be asymptotically flat and stationary.

The paper is organized as follows. In Section~\ref{sec:Bohm} we recall the relevant version of Bohm's law of motion. In Section~\ref{sec:RN} we recall the Reissner--Nordstr\"om geometry. In Section~\ref{sec:singular} we discuss the general framework of particle creation and annihilation at a timelike singularity. In Section~\ref{sec:concrete} we describe the concrete model. In Section~\ref{sec:BTQFT} we draw parallels between our model and Bell-type quantum field theories. In Section~\ref{sec:foliation} we explain why foliations defined by constant lapse function cannot be used here. In Section~\ref{sec:conclusions} we conclude.

\section{Bohmian Mechanics}
\label{sec:Bohm}

\subsection{Nature of the Theory}

Bohmian mechanics  \cite{Bohm52,Bell87b,BH,DGZ92} is a candidate for how quantum phenomena work; see \cite{Gol01,Bri16} for introduction and overview. While the quantum formalism describes what observers will see, and while the conventional ``Copenhagen'' view of quantum mechanics remains vague and paradoxical, Bohmian mechanics provides a possible explanation of the quantum formalism in terms of objective events, in fact by postulating that particles have actual positions and hence trajectories. Bohmian mechanics is well understood in the realm of non-relativistic quantum mechanics, but needs further development in the directions of relativistic physics, quantum field theory, and quantum gravity. This paper concerns the relativistic extension in a classical gravitational field, but connects also with quantum field theory.

Bohmian mechanics postulates that particles have trajectories, governed by an equation of motion of the type
\begin{equation}\label{Bohm}
  \frac{dQ_t}{dt} = \frac{J_{\psi_t}(Q_t)}{\rho_{\psi_t}(Q_t)}\,,
\end{equation}
where $Q_t$ is the position of the particle at time $t$ (or, for a system of several particles, the \emph{configuration}), and $J_\psi$ and $\rho_\psi$ are, respectively, the quantum mechanical probability current and probability density ($|\psi|^2$ or a variant thereof) as determined by the wave function $\psi$. For example \cite{Bohm53}, if $\psi$ is a Dirac wave function of a single particle then
\begin{equation}\label{jDirac}
  J_\psi = \psi^\dagger \alpha \psi\,, \quad \rho_\psi = \psi^\dagger \psi
\end{equation}
with $\alpha=(\alpha^1,\alpha^2,\alpha^3)$ the three Dirac alpha matrices. Equivalently, the possible world lines of the Bohmian particle are the integral curves of the current vector field 
\be
j^\mu = \overline{\psi} \gamma^\mu \psi\,.
\ee
As a consequence of the structure \eqref{Bohm} of the law of motion, if at any time $t$ the particle position (or configuration) is random with distribution $\rho_{\psi_t}$, then this is also true of any other time $t$. This property is called \emph{equivariance}. As a (non-obvious) consequence of \emph{that}, inhabitants of a Bohmian universe, consisting of these particles with trajectories, would observe the same probabilities in their experiments as predicted by the quantum formalism \cite{DGZ92}. That is how Bohmian mechanics explains quantum mechanics. In fact, Bohmian mechanics accounts for all phenomena of non-relativistic quantum mechanics.

The question whether Bohmian trajectories might run into a singularity has already been considered in a different context, that of the non-relativistic theory. Here, ``singularity'' does not refer to a space-time singularity (i.e., points of infinite curvature), but to singularities of the potential, of the wave function, or of the velocity vector field. For example, for a Bohmian particle moving in a Coulomb potential $1/|\boldsymbol{x}|$, one may worry about trajectories running into the origin, where the potential is singular. Likewise, one may worry about trajectories running into points where $\psi$ is non-differentiable (in the non-relativistic case) or zero, which is where the velocity vector field is not defined. It turns out \cite{ex,TT05} that with probability 1, the trajectory never runs into such points, and this has to do with the vanishing probability current into such points. The situation is different in our setting, where Bohmian trajectories do run into the space-time singularity with positive probability, corresponding to the nonzero current into the singular set. On the other hand, our situation is quite analogous to, and can be regarded as a special or limiting case of, Bohmian motion in a manifold with boundary, which suggests the use of interior-boundary conditions \cite{bohmibc}.

\subsection{Bohmian Mechanics and Relativity}

It is known \cite{HBD,timbook,3forms} that, if we are willing to consider a preferred foliation $\foliation$ of space-time into spacelike 3-surfaces, which I will call the \emph{time foliation}, then Bohmian mechanics possesses a natural generalization to relativistic space-time. The possibility of a preferred foliation seems against the spirit of relativity (see \cite{timbook} for a discussion), but certainly worth exploring. It is suggested by the empirical fact of quantum non-locality, and it is suggested by the structure of the Bohmian law of motion \eqref{Bohm} for many particles, in which the velocity of a particle depends on the instantaneous position of the other particles. The time foliation could be given by a Lorentz invariant law \cite{Tum06d,DGNSZ14}.

For a time foliation $\foliation$ (determined in whichever way), a Bohm-type equation of motion was formulated in \cite{HBD} for flat space-time (based on earlier work in \cite{Bohm53,BH}); the straightforward generalization to curved space-time \cite{3forms} reads:
\begin{equation}\label{hbd}
  \frac{dX_k^{\mu_k}}{d\tau} \propto j^{\mu_1 \ldots \mu_N} 
  \bigl(X_1(\Sigma),\ldots, X_N(\Sigma)\bigr) 
  \prod_{i\neq k} n_{\mu_i}\bigl(X_i(\Sigma)\bigr)\,,
\end{equation}
where $X_k(\tau)$ is the world line of particle $k\in \{1,\ldots,N\}$, $\tau$ is any curve parameter, $\Sigma$ is the 3-surface in $\foliation$ containing $X_k(\tau)$, $n(x)$ is the unit normal vector on $\Sigma$ at $x \in \Sigma$, $X_i(\Sigma)$ is the point where the world line of particle $i$ crosses $\Sigma$, and
\begin{equation}\label{multij}
  j^{\mu_1 \ldots \mu_N} = \overline{\psi} (\gamma^{\mu_1} \otimes \cdots \otimes
  \gamma^{\mu_N}) \psi
\end{equation}
is the probability current tensor associated with the $N$-particle Dirac wave function $\psi$. This wave function could either be a multi-time wave function defined on the pairwise-spacelike elements of (space-time)$^N$ \cite{PT13a,PT13b,Lie15,LN15,LPT17}, or, since we never use $\psi$ for configurations that are not simultaneous relative to $\foliation$, it suffices that $\psi$ is defined on the $3N+1$-dimensional manifold $\bigcup_{\Sigma \in \foliation} \Sigma^N$ of simultaneous configurations.

This generalization does not automatically include, however, space-time geometries with singularities. The treatment of singularities requires, as we point out in Section~\ref{sec:singular} below, some fundamental extensions of Bohmian mechanics, and forms a test case for the robustness of the equation of motion \eqref{hbd}. 

As mentioned before, the foliation might itself be dynamical, i.e., might be determined through a law by initial data. An example of a possible Lorentz invariant evolution law for the foliation is
\begin{equation}\label{lawF}
  \nabla_{\!\mu} n_\nu - \nabla_{\!\nu} n_\mu =0\,,
\end{equation}
or, in the language of differential forms,
\be\label{dn=0}
dn=0\,,
\ee
which is equivalent to saying that the infinitesimal timelike distance between two nearby 3-surfaces from the foliation is constant along the 3-surface, or that the lapse function is constant \cite{Tum06d}. This law allows to choose an initial spacelike 3-surface and then determines the foliation. A special foliation obeying \eqref{lawF} is the one consisting of the surfaces of constant timelike distance from the Big Bang. Note, however, that the law of motion \eqref{hbd} does not require any particular choice of law for the foliation, except that the foliation does not depend on the particle configuration (while it may depend on the wave function). It has been found that the equation of motion \eqref{hbd} still makes sense (and leads to equivariance) even when the leaves of the time foliation have kinks \cite{ST14} or overlap \cite{ST15}. 
Note further that in a space-time with horizons, a foliation law such as \eqref{lawF} will frequently lead to 3-surfaces lying partly outside and partly inside the horizon. 

It will turn out from our study of  singularities that \eqref{lawF} is not a good law for the time foliation in the presence of a timelike singularity as the foliations obeying \eqref{lawF} may fail to cover all of space-time; see Section~\ref{sec:foliation} for details.

\section{Reissner--Nordstr\"om Geometry}
\label{sec:RN}

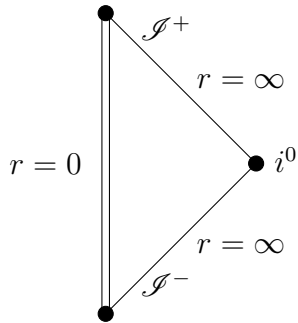
\begin{figure}[h]
\begin{center} 
  \begin{tikzpicture}
  \draw (-0.05,2) -- (-0.05,-2);
  \draw (0.05,2) -- (0.05,-2);
  \draw (0,2) -- (2,0) -- (0,-2);
  \filldraw (0,-2) circle [radius=0.1];
  \filldraw (0,2) circle [radius=0.1];
  \filldraw (2,0) circle [radius=0.1];
  \node at (2.4,0.05) {$i^0$};
  \node at (-0.8,0) {$r=0$};
  \node at (0.8,1.8) {$\mathscr{I}^+$};
  \node at (1.8,1.1) {$r=\infty$};
  \node at (0.8,-1.6) {$\mathscr{I}^-$};
  \node at (1.8,-1.1) {$r=\infty$};
  \end{tikzpicture}
\end{center}
 \caption{Penrose conformal diagram \cite{Pen64,HE73} of the super-critical Reissner--Nordstr\"om space-time. The double line on the left marks the singularity.}\label{fig:Penrose}
\end{figure}

Our main example of a timelike singularity will be the one of the \emph{Reissner--Nordstr\"om} space-time geometry \cite{Nor13,rei16,wey18,Nor18,MTW,HE73}, a solution of the coupled Einstein and Maxwell equations for a charged point mass. It is given by
\be\label{RN}
ds^2 = \lambda(r) dt^2 -
\frac{1}{\lambda(r)} dr^2 -
r^2 (d\vartheta^2 + \sin^2 \vartheta \, d\varphi^2)
\ee
with 
\be
\lambda(r) = 1-\frac{2M}{r} + \frac{e^2}{r^2} 
\ee
and parameters $M>0$ and $e\in \RRR$. The Reissner--Nordstr\"om metric is spherically symmetric, static, asymptotically flat, and singular at $r=0$. It arises as the gravitational field of a (non-rotating) point particle with mass $M$ and charge $e$. We assume the so-called \emph{super-critical} case $|e|>M$; in this case $\lambda(r)>0$ for all $r\geq 0$. The Penrose diagram is shown in Figure~\ref{fig:Penrose}. In this context we note that if an electron were a classical point particle, its gravitational field would be a super-critical Reissner--Nordstr\"om field containing a timelike singularity. This does not mean that electrons must, but does suggest they might, contain timelike singularities in nature. 

We find it useful to regard the singularity $\{r=0\}$ not as a \emph{line} (as suggested by the picture of a point particle, and by the fact that in Minkowski space-time $\{r=0\}$ is a line) but as a \emph{surface} (of topology $\RRR\times \SSS^2$). The main reason for this approach is that the boundary condition on the wave function will depend on the direction from which the singularity is approached, but this approach is also natural in spherical coordinates, in which $\{r=0\}$ is a surface of codimension 1. We thus regard the space-time
\be
\st \cong \bigl\{(t,r,\omega): t\in \RRR, r\geq 0, \omega \in \SSS^2\bigr\} 
= \RRR\times [0,\infty) \times \SSS^2
\ee
(where $\cong$ means diffeomorphic) as a manifold with boundary. (For the mathematical notion of a manifold with boundary, see \cite{Lan72,BM}. We will sometimes write $\omega$ for a point on $\SSS^2$, and sometimes use the standard parameterization $(\vartheta,\varphi)$ by polar and azimutal angle.) The singularity is the \emph{boundary}
\be
\sing =\{r=0\}=\boun \cong \RRR\times \{0\}\times \SSS^2\,,
\ee
while the metric $g_{\mu\nu}$ is defined on the \emph{interior} $\inte \cong \RRR\times (0,\infty)\times\SSS^2$. 

As the \emph{time foliation} $\foliation$ for defining the Bohmian trajectories we use the foliation provided by the time coordinate $t$, i.e., the family of level surfaces of the $t$ function:
\be
\foliation = \bigl\{ \Sigma_s: s\in\RRR  \bigr\} \quad
\text{with}\quad \Sigma_s = \{x\in \st:t(x)=s\}\,. 
\ee
Concretely, $\Sigma_s = \{s\}\times [0,\infty) \times \SSS^2$. The fact that we are considering a static space-time invites us to regard $\Sigma := [0,\infty)\times \SSS^2$ as ``the'' 3-space.

\section{Jumps in Configuration Space}
\label{sec:singular}

Bohmian trajectories, the integral curves of the probability current, are \emph{causal} curves, i.e., their tangent vectors are always timelike or lightlike (when based on the Dirac equation). Such a curve may well reach the singularity $\boun$ (indeed, in finite coordinate time $t$ and finite proper time). What should happen if we consider a system of $n$ particles and one of them hits the singularity? The simplest possibility, or at least a natural possibility, is that this particle gets annihilated, that it stops existing, and history proceeds with $n-1$ particles. This is the possibility we study in this paper.

In Bohmian theories, a configuration $q$ of $n$ identical particles is described by a set of $n$ points in 3-space; thus, if $\Sigma$ is 3-space, then the configuration space is
\be\label{Qndef}
\conf_n = \{q\subset \Sigma: \# q = n\}\,,
\ee
the space of unordered configurations in $\Sigma$.
The wave function determining the motion of $n$ particles is best regarded as a function on $\widehat\conf_n$ \cite{topid1A}, the universal covering space \cite{Cov} of $\conf_n$ or, equivalently, $\widehat\conf_n = \Sigma^n$, the space of ordered configurations in $\Sigma$.\footnote{Actually, the covering space of $\conf_n$ is $\{(x_1,\ldots,x_n)\in\Sigma^n:x_i\neq x_j \forall i\neq j\}$, the set of collision-free ordered configurations. However, for our considerations it will not matter whether we include or exclude the collision configurations.} If the number of particles can change then we need several wave functions, one for each particle number. This leads us to consider state vectors in \emph{Fock space}
\be
\Hilbert = \bigoplus_{n=0}^\infty \Hilbert_n\,,
\ee
where $\Hilbert_n$ is the Hilbert space of $n$-particle wave functions, $\Hilbert_n = S_\pm \Hilbert_1^{\otimes n}$, where $S_+$ means the symmetrizer (appropriate for bosons), $S_-$ the anti-symmetrizer (for fermions), and $\Hilbert_1$ the one-particle Hilbert space comprising the square-integrable spinor-valued functions on $\Sigma$ (more detail later). Every element $\psi \in \Hilbert$ can thus be regarded as a sequence
\be
\psi=(\psi^{(0)},\psi^{(1)},\ldots,\psi^{(n)},\ldots)\,,
\ee
where $\psi^{(n)}$ is an $n$-particle wave function, defined on $\Sigma^n$. Corresponding to the structure of Fock space, the configuration space is
\be\label{confdef}
\conf = \bigcup_{n=0}^\infty \conf_n\,,
\ee
where $\conf_n$ is the configuration space of $n$ particles, as defined in \eqref{Qndef}.
The Fock state $\psi$ can then be regarded as a function on the covering space
\be
\widehat\conf = \bigcup_{n=0}^\infty \widehat\conf_n\,.
\ee
(Both $\conf_0$ and $\widehat\conf_0$ contain just one element, the empty configuration.) 

In the theory we are developing, if one among $n$ particles hits the singularity, the configuration jumps from $\conf_n$ to $\conf_{n-1}$ in a deterministic way: $q \to \dest(q)$, where the jump function $\dest: \partial \conf_n \to \cup_{m<n}\conf_{m}$ is given by 
\be\label{destdef}
\dest(q) = q \setminus \partial \Sigma\,.
\ee
Here, $\partial \Sigma$, the boundary of $\Sigma$, is the intersection of $\Sigma$ with the singularity $\sing =\boun$; $\partial \conf_n$ is the set of configurations in which at least one particle is located on $\partial \Sigma$; and \eqref{destdef} means that $f$ just removes all particles located on $\partial \Sigma$, i.e., all particles that have arrived at the singularity.

The situation in which the configuration space is a manifold with boundaries has been considered before in \cite{GT05b,bohmibc}; we recapitulate the Bohmian dynamics of the configuration developed there. It includes deterministic jumps $Q\to f(Q)$ that occur whenever the configuration $Q$ reaches the boundary $\partial \conf$ of configuration space, according to some given $\dest$. We desire three properties of the theory: time reversal invariance, equivariance, and the Markov property; as explained in \cite{GT05b,bohmibc}, these properties entail stochastic jumps from configurations in the interior $\interior=\conf\setminus \boundary$ to the boundary $\boundary$ with a certain rate. 
This leads us to a stochastic process $(Q_t)_{t\in \RRR}$ in $\conf$
consisting of smooth motion interrupted by jumps from the boundary to
the interior $\interior$ or vice versa.  

The deterministic jump law, abstractly
\be\label{detjump}
      Q_{t+} = \dest(Q_{t-})
\ee
at any time $t$ at which the trajectory hits the boundary, is based on the given, fixed mapping $\dest:\boundary \to \interior$, in our case given by \eqref{destdef}.
Since we want the theory to be reversible, we must also allow for
spontaneous jumps from interior points to boundary points.  Since we
want the process to be an equivariant Markov process, the rate for a
jump from $q' \in \interior$ to a surface element $\D q \subseteq
\boundary$ must be, as one can derive \cite{bohmibc},
\begin{equation}\label{boundrate}
      \sigma_t(q'\to \D q) = \frac{ J^\perp_{\psi_t}(q)^+} 
      {\rho_{\psi_t}(q')} \, \antidest(\D q,q'),
\end{equation}
where $s^+ = \max(s,0)$ denotes the positive part of $s\in\RRR$, $J_\psi$ is the probability current in configuration space relative to coordinate volume (see Section~\ref{sec:current} below), $J_\psi^\perp$ is its component orthogonal to $\boundary$ (in our case,
the radial component), $\rho_\psi$ is the probability density relative to coordinate volume, and $\antidest(\D q,q')$ is the uniform measure on $f^{-1}(q')$; more precisely, it is the measure-valued function defined in terms of the volume measure $\vol$ on $\conf$ and the surface area measure $\area$ on $\boundary$ by
\be
 \antidest(B,q') = \frac{\area(B \cap \dest^{-1}(\D q'))}{\vol(\D q')},
\ee
for every $B \subseteq \boundary$. In our case, the measure $\vol$ is on $\conf_n$
locally the product of $n$ copies of the coordinate volume measure on 
$\Sigma = [0,\infty)\times\SSS^2$, and the measure $\lambda$ is on $\boundary_n$ locally the product of the surface area measure on $\SSS^2$ and $n-1$ copies of the coordinate volume measure on $\Sigma = [0,\infty)\times\SSS^2$. Note that the set of configurations with two or more particles on the singularity has $\lambda$-measure zero. It follows that $\antidest(\cdot,q')$ is a measure concentrated on the set of those configurations that consist of $q'$ plus one further particle on the singularity $\partial \Sigma\cong \SSS^2$, and the measure is essentially a copy of the surface area measure on $\SSS^2$. Thus, concretely, the creation rate density relative to the area measure on $\SSS^2$ is 
\begin{equation}\label{ratedensity}
      \sigma_t\bigl(q\to q\cup \omega\bigr) = \frac{ J_{\psi_t}^\perp(q\cup\omega)^+} 
      {\rho_{\psi_t}(q)} \,,
\end{equation}
where we have simply written $\omega$ for the point $(r=0,\omega)$ on the singularity $\sing \cap\Sigma$.

Note that a jump to a boundary point $q$ at
which the current is pointing \emph{towards} the boundary,
$J^\perp(q) < 0$, would not allow any continuation of the
process since there is no trajectory starting from $q$. The problem is
absent if the velocity at $q$ is pointing \emph{away} from the
boundary, $J^\perp(q) >0$. (We ignore the case $J^\perp(q) = 0$.) On the other hand, jumps from $q$ to $\dest(q)$ cannot
occur when $J(q)$ is pointing away from the boundary since in that
case there is no trajectory arriving at $q$. Thus, the jumps must be
such that at each time $t$, one of the transitions $q \to \dest(q)$ or
$\dest(q) \to q$ is forbidden, and which one it is depends on the sign of
$J^\perp_{\psi_t}(q)$.

Given the law of motion
\be\label{Bohm2}
\frac{dQ_t}{dt} = v(Q_t) = \frac{J}{\rho}(Q_t)
\ee
together with the deterministic jump law \eqref{detjump} and stochastic jumps with rate \eqref{boundrate}, we obtain the following transport equation for the probability density $p$ at $q' \in \interior$:
\begin{equation}\label{continuity2}
      \frac{\partial p}{\partial t}(q',t) = -\sum_{i=1}^{3n}\partial_i 
  \bigl(v^i p\bigr)\big|_{(q',t)} -\sigma_t(\boundary,q') \, p(q',t) + \!
      \int\limits_{\boundary} \!\antidest(\D q,q') \,v^\perp(q,t)^- p(q,t) 
\end{equation}
with $s^-=\max(-s,0)$ the negative part of $s\in\RRR$, $s=s^+-s^-$.
For equivariance, we need that whenever $p=\rho_{\psi_t}$ then $\partial_t p= \partial_t \rho_{\psi_t}$. This follows if \eqref{continuity2} with $\rho_{\psi_t}$ inserted for $p$ agrees with the continuity equation for $\rho_{\psi_t}$ that follows from the Schr\"odinger equation $i\hbar\partial_t \psi_t = H \psi_t$. This is what our model achieves by means of a suitably modified Dirac evolution and an interior--boundary condition at the singularity.

\section{Concrete Model}
\label{sec:concrete}

For our model, it will be convenient to consider spin-$\tfrac{1}{2}$ bosons (even though spin-$\tfrac{1}{2}$ particles are fermions in nature). 

\subsection{One-Particle Dirac Equation}

There is a standard way \cite{LM89,PR84} of setting up the one-particle Dirac equation in a curved space-time; in a space-time without singularities (more precisely, in a globally hyperbolic space-time), the Dirac equation will by itself (i.e., without any boundary conditions) define the time evolution, i.e., will determine $\psi$ on all of $\st$ from initial data on a Cauchy surface. This standard setup (for a space-time with unique spin structure, such as the super-critical Reissner-Nordstr\"om space-time \cite[Sec.~1.5]{PR84}) comprises the construction of a bundle $\spin= \cup_{x\in\st}\spin_x$ of Dirac spin spaces that are 4-dimensional complex vector spaces (of which $\psi$ will be a section); of an indefinite sesquilinear form on $\spin_x$ written as $\overline\phi \psi$; of a section $\gamma$ of $\CCC T\st \otimes \spin \otimes \spin^*$ (where $\CCC T\st$ is the complexified tangent bundle, $\otimes$ the fiber-wise complex tensor product, and $\spin^*$ the bundle of dual spaces of $\spin_x$) satisfying the Clifford relation
\be\label{Clifford}
\gamma^\mu \gamma^\nu + \gamma^\nu \gamma^\mu = 2 g^{\mu\nu} I
\ee
with $I$ the identity mapping of $\spin_x$; and of a connection $\nabla$ on $\spin$ determined by the metric $g$ on $\st$. The one-particle Dirac equation on $\st$ then simply reads
\be\label{Dirac}
i\hbar\gamma^\mu \nabla_\mu \psi = m\psi\,.
\ee
The one-particle Hilbert space $\Hilbert_1$ associated with the spacelike hypersurface $\Sigma$ (we consider only $\Sigma=\{t=t_0\}$ and can take $t_0=0$ since the Reissner--Nordstr\"om space-time is static) is equipped with the inner product
\be
\langle \phi | \psi \rangle = \int_\Sigma d^3x \, \overline{\phi}(x)\, \gamma^\mu(x)\, n_\mu(x) \, \psi(x)
\ee
with $d^3x$ the volume associated with the 3-metric on $\Sigma$ and $n^\mu(x)$ the future unit normal vector to $\Sigma$ at $x$. It is common to exclude wave functions of negative energy as unphysical, but we will not exclude them in our model.

\subsection{Probability Current in Space-Time}
\label{sec:current}

When we speak of probability density, or probability current density, it plays a role that densities can be considered either relative to \emph{coordinate volume} or relative to \emph{invariant volume} (proper volume), and this leads us to consider two current vector fields on $\st$ (and afterwards on $\conf$), which we denote respectively $\tilde{\jmath}^\mu$ and $j^\mu$. We elucidate this below in more detail, beginning with the single-particle case. The reason we consider densities relative to coordinate volume is that they often stay bounded at the singularity (if the coordinates are suitably chosen), whereas the relevant densities relative to invariant volume often diverge.

The current given by a one-particle wave function, $j^\mu = \overline{\psi}\gamma^\mu \psi$, is the current relative to invariant volume, which means the following. For every spacelike hypersurface $\Sigma\subseteq \st$, let $n^\mu(x,\Sigma)$ be the future-pointing unit normal vector on $\Sigma$ at $x\in\Sigma$; let $\lambda_\Sigma(d^3x)$ be the 3-volume measure associated with the Riemannian 3-metric on $\Sigma$ inherited from $g_{\mu\nu}$. The Bohmian trajectory associated with $\psi$ is a randomly chosen integral curve $L$ of $j^\mu$ whose probability distribution $\PPP$ is such that, for every spacelike hypersurface $\Sigma$, the trajectory's unique intersection point with $\Sigma$ has distribution $j^\mu \, n_\mu \, \lambda_\Sigma$, i.e.,
\be\label{jdist}
\PPP\bigl(L\cap\Sigma \subseteq A\bigr) = \int_A j^\mu(x)\,g_{\mu\nu}(x)\, n^\nu(x,\Sigma) \, \lambda_\Sigma(d^3x)\,,
\ee
where $A$ is any subset of $\Sigma$, $L\cap \Sigma$ the (set containing only the) intersection point, and $\PPP(L\cap\Sigma \subseteq A)$ the probability of this point lying in $A$.

When computing densities relative to coordinate volume, we regard the coordinate space, in our example $\RRR\times[0,\infty)\times\SSS^2$, as a Riemannian 4-manifold $\tilde\st$ with (positive definite) metric $\tilde{g}_{\mu\nu}$, in our example
\be\label{tildegdef}
d\tilde{s}^2 = dt^2 + dr^2 + d\vartheta^2 + \sin^2 \vartheta\, d\varphi^2\,.
\ee
(Note that this is a little different from what would usually be regarded as the coordinate space, which is $\RRR\times [0,\infty)\times [0,\pi) \times [0,2\pi)$ with metric $dt^2 + dr^2 + d\vartheta^2 + d\varphi^2$. The reason we use a different space is to avoid dealing with the coordinate singularities of $\vartheta$ and $\varphi$.)

Every spacelike hypersurface $\Sigma\subseteq \st$ can also be regarded as a hypersurface in $\tilde\st$; let $\tilde{n}^\mu(x,\Sigma)$ be the vector at $x\in\Sigma$ that is normal in coordinates (i.e., in $\tilde{g}_{\mu\nu}$) to $\Sigma$, has unit length in coordinates, and points to the same side of $\Sigma$ as $n^\mu(x,\Sigma)$. Likewise, let $\tilde{\lambda}_\Sigma(d^3x)$ be the 3-volume measure defined by the 3-metric inherited from $\tilde{g}_{\mu\nu}$. As we will show presently, there is a unique vector field $\tilde{\jmath}^\mu$ such that, for every spacelike hypersurface $\Sigma$,
\be\label{tildej}
\tilde{\jmath}^\mu(x) \, \tilde{g}_{\mu\nu} \, \tilde{n}^\nu(x,\Sigma) \, \tilde{\lambda}_\Sigma(d^3x) = j^\mu(x)\,g_{\mu\nu}(x)\, n^\nu(x,\Sigma) \, \lambda_\Sigma(d^3x)\,.
\ee
Indeed,
\be\label{jtildej}
\tilde{\jmath}^\mu(x) = \frac{d\varepsilon}{d\tilde\varepsilon}(x) \, j^\mu(x)\,,
\ee
where $\varepsilon(d^4x)$ is the invariant 4-volume measure associated with $g_{\mu\nu}$, $\tilde\varepsilon(d^4x)$ the coordinate 4-volume measure (associated with $\tilde{g}_{\mu\nu}$), and $d\varepsilon/d\tilde\varepsilon$ the density of one relative to the other (also known as the Radon--Nikodym derivative). Note that, as a consequence of the proportionality, $\tilde{\jmath}^\mu$ and $j^\mu$ have the same integral curves (up to an irrelevant reparameterization).

To verify \eqref{tildej} and \eqref{jtildej}, we use differential forms (see, e.g., \cite{MTW}). Let $\boldsymbol{\varepsilon}$ (in bold face) be the invariant volume 4-form associated with $g_{\mu\nu}$, and let $\boldsymbol{j}$ be the 3-form obtained by inserting $j^\mu$ into the first slot of $\boldsymbol{\varepsilon}$. Then \cite{3forms} the distribution \eqref{jdist} can be written as
\be
\PPP\bigl(L\cap\Sigma \subseteq A\bigr) = \int_A \boldsymbol{j}
\ee
for every subset $A$ of $\Sigma$. That is, the 3-form $\boldsymbol{j}$ encodes the probability current without reference to any metric. Making the same step backwards with $\tilde{g}_{\mu\nu}$, we obtain that
\be
\tilde{\jmath}^\mu \, \boldsymbol{\tilde{\varepsilon}}_{\mu\nu\kappa\lambda} = \boldsymbol{j}_{\nu\kappa\lambda} = j^\mu \, \boldsymbol\varepsilon_{\mu\nu\kappa\lambda}\,, 
\ee
which implies \eqref{tildej} and \eqref{jtildej}, as $\boldsymbol\varepsilon_{\mu\nu\kappa\lambda}(x) = \frac{d\varepsilon}{d\tilde\varepsilon}(x) \, \boldsymbol{\tilde{\varepsilon}}_{\mu\nu\kappa\lambda}(x)$.

One can compute the Radon--Nikodym derivative between the two measures by 
\be
\frac{d\varepsilon}{d\tilde\varepsilon} = 
\frac{\sqrt{-\det (g_{\mu\nu})}}{\sqrt{\det (\tilde{g}_{\mu\nu})}}\,.
\ee
In our example case,
\be
\frac{d\varepsilon}{d\tilde\varepsilon} = r^2\,.
\ee 
(Note that taking $\RRR\times (0,\infty)\times\SSS^2$ as coordinate space differs by a factor of $r^2$ from taking $\RRR\times(\RRR^3\setminus\{0\})$, which would have come with a further factor of $r^2$ in front of $d\vartheta^2+\sin^2 \vartheta \, d\varphi^2$ in \eqref{tildegdef}.) We set
\be
\tilde{\gamma}^\mu(x) = \frac{d\varepsilon}{d\tilde\varepsilon}(x) \, \gamma^\mu(x)\,,
\ee
so that $\tilde{\jmath}^\mu = \overline{\psi}\tilde{\gamma}^\mu\psi$.

If $\Sigma$ is not spacelike, the distribution of the intersection point $L \cap \Sigma$ is still given by
\be\label{tildejdist}
\PPP\bigl(L\cap\Sigma \subseteq A\bigr) = \int_A \tilde{\jmath}^\mu(x)\,\tilde{g}_{\mu\nu}(x)\, \tilde{n}^\nu(x,\Sigma) \, \tilde{\lambda}_\Sigma(d^3x)
\ee
for any $A\subseteq \Sigma$, provided that $L$ intersects $\Sigma$ at most once, and that $\tilde{n}^\nu$ always points to the same side of $\Sigma$ as $\tilde{\jmath}^\mu$. In particular, if $\Sigma=\sing $ is the singularity, then we obtain that the probability distribution of the random point where $L$ hits the singularity has density (in coordinates) equal to (the absolute value of) $\tilde{\jmath}^r$, the radial component of $\tilde{\jmath}^\mu$, except where $\tilde{\jmath}^r>0$ so that there is a current \emph{away from} the singularity.

Since we only need that $\tilde{\jmath}^\mu$ remains bounded at the singularity, we can allow $j^\mu$ to diverge like $r^{-2}$. This means, since $j^\mu = \overline{\psi} \gamma^\mu \psi$, that the components of $\psi$ (relative to chosen coordinates and bases in $\spin_x$) can diverge like $r^{-1}$ while $\gamma^\mu$ stays bounded, or those of $\gamma^\mu$ can diverge like $r^{-2}$ while $\psi$ stays bounded (or other possibilities). We will choose the second option, so that relevant wave functions $\psi$ remain bounded at the singularity. This is connected to the definition of the spin spaces at the singularity.

\subsection{Spin Spaces at the Singularity}

The spin spaces $\spin_x$ are standardly defined at interior points $x\in\inte=\st\setminus \sing $, and we also need to define them at points $x\in \sing $ in order to be able to talk of $\psi(x)$ and $\tilde{\jmath}^\mu(x)$ for $x\in \sing $. As a preparation, let us look at how the tangent spaces $T_x\st$ are defined for $x\in \sing $. (We note that, according to standard definitions of the tangent space, $T_x\st$ is a vector space---not a half-space.) Using coordinates, we can simply say, for $x=(t,r=0,\omega)$, that
\be
T_x\st = \RRR\times \RRR\times T_\omega \SSS^2\,,
\ee
where the first factor $\RRR$ is understood as the tangent space to the manifold $\RRR$ at $t$, and the second as the tangent space to the manifold-with-boundary $[0,\infty)$ at $0$. 

As a part of this specification of tangent spaces, we intend a certain topology (and differentiable structure) on the bundle $T\st= \cup_{x\in \st} T_x \st$: for example, if we set $x=(t,r,\omega), y=(t,0,\omega)$, keep $t$ and $\omega \in\SSS^2$ fixed and let $r\to 0$ so that $x\to y$, then the unit normal to the level surface $\Sigma_t$ of the $t$ function, $n^\mu(x)$, converges to $0$, as its coordinate components are $(\lambda(r)^{-1/2},0,0)$, while the unit normal relative to $\tilde{g}_{\mu\nu}$, $\tilde{n}^\mu(x)$, converges to $\tilde{n}^\mu(y)$. Relatedly, another way of defining the tangent spaces on the singularity consists in specifying a tetrad field in $\inte$ (i.e., a basis for every tangent space) that possesses a smooth extension to $\boun$. In our example, we can take as the tetrads in $\inte$ simply the coordinate bases (or, orthonormal bases relative to $\tilde{g}_{\mu\nu}$). Let me emphasize that the bundle $T\st$ over the space-time with boundary is not determined by the bundle $T\inte$ over the interior of space-time; there are inequivalent possibilities for the extension of the bundle to $\sing =\boun$, and we have chosen one by demanding that our coordinate basis field in $T\st$ forms a smooth extension of our coordinate basis field in $T\inte$.

In the same way, we define the spin space $\spin_x$ for $x\in\boun$ by specifying a basis $\tilde{b}_x$ of $\spin_x$ for every $x\in\inte$ and postulating that the field $\tilde{b}_x$ possesses a smooth extension to $\boun$. To this end, fix any $x\in\inte$ and begin with the coordinate basis $\tilde{e}_x$ of $T_x\st$, $\tilde{e}_x=(\partial_t,\partial_r, \partial_\vartheta,\partial_\varphi)$. By normalizing these four vectors, we transform them into an orthonormal basis (i.e., Lorentz frame, also known as Minkowski tetrad) $e_x= (\lambda^{-1/2}\partial_t, \lambda^{1/2}\partial_r,r^{-1}\partial_\vartheta, (r\sin \vartheta)^{-1}\partial_\varphi)$. To this orthonormal basis there corresponds a basis\footnote{The same can be done in two-spinor calculus, where this basis is called a ``spin frame'' \cite{PR84}.} $b_x$ of $\spin_x$; the correspondence is canonical up to an overall sign which we choose continuously in $x$; $b_x$ is an orthonormal basis relative to the scalar product $\overline{\phi} \, \gamma^\mu(x) \, g_{\mu\nu}(x)\, n^\nu(x) \, \psi$ in $\spin_x$ (note that $n(x) = \lambda^{-1/2} \partial_t$). This basis we rescale by $\lambda(r)^{-1/4} \,(d\varepsilon/d\tilde\varepsilon)^{-1/2}$ to obtain $\tilde{b}_x$. This completes the definition of $\tilde{b}_x$.

In other words, if we define the sesquilinear mappings $\alpha,\tilde{\alpha}:\spin_x \times \spin_x \to \CCC T_x\st$ by
\be
\alpha(\phi,\psi) =  \overline{\phi} \gamma^\mu \psi\,,\quad
\tilde\alpha(\phi,\psi) =  \overline{\phi} \tilde\gamma^\mu \psi\,,
\ee
then the coefficients $\alpha^{\mu}_{~ss'}$ of $\alpha$ relative to $b_x$ and $e_x$ are the usual three Dirac $\alpha$ matrices, in the following denoted
\be
\underset{0}{\alpha}^i= \begin{pmatrix}0 & \sigma^i\\ \sigma^i & 0 \end{pmatrix},\text{ together with }\underset{0}{\alpha}^0=I
\ee
(where $\sigma^1,\sigma^2,\sigma^3$ are the Pauli matrices and $I$ is the identity matrix), while the coefficients of $\tilde{\alpha}$ relative to $\tilde{b}_x$ and $\tilde{e}_x$ are as follows:
\be\label{tildealpha}
(\tilde\alpha^t,\tilde\alpha^r,\tilde\alpha^\vartheta,\tilde\alpha^\varphi) =
\Bigl( \frac{1}{\lambda}I, \, \underset{0}{\alpha}^1, \frac{1}{\lambda^{1/2}r} \, \underset{0}{\alpha}^2, \frac{1}{\lambda^{1/2}r\sin\vartheta} \, \underset{0}{\alpha}^3 \Bigr)\,.
\ee
This equation could also be taken as the definition of $\tilde{b}_x$. We define the differentiable bundle structure of the spin bundle $\spin$ over $\st$ so that $\tilde b_x$ depends smoothly on $x\in\st$, also on the boundary $\boun$. 

It follows that the coordinate formula for the probability current is given by
\be
\tilde{\jmath}^\mu = \psi^* \tilde{\alpha}^\mu \psi
\ee
where the components of $\psi$ are taken relative to the basis $\tilde{b}_x$, and the $\tilde{\alpha}^\mu$ matrices are given by the right hand side of \eqref{tildealpha}. 
We note for later use that these matrices possess the following limits
\be\label{tildealphaS}
\text{as }r\to0\text{:}\quad\quad(\tilde\alpha^t,\tilde\alpha^r,\tilde\alpha^\vartheta,\tilde\alpha^\varphi) \to
\Bigl( 0, \, \underset{0}{\alpha}^1, \frac{1}{e} \, \underset{0}{\alpha}^2, \frac{1}{e\sin\vartheta} \, \underset{0}{\alpha}^3 \Bigr)\,,
\ee
with $e$ the charge parameter in \eqref{RN}. 

In the following, we will express wave functions relative to the basis $\tilde{b}_x$ of spin space. If we considered only a single particle without particle creation and annihilation, the Schr\"odinger equation would read
\begin{align}
i\hbar \, \partial_t \psi(x) 
&=   -i\hbar \sum_{i=1}^3 \tilde\alpha^t(x)^{-1}\tilde\alpha^i(x) \, \partial_{i} \psi(x) + V(x) \, \psi(x) \\
&=   -i\hbar \sum_{i=1}^3 \lambda(x) \, \underset{0}{\alpha}^i \, \partial_{i} \psi(x) + V(x) \, \psi(x) 
\end{align}
in coordinates $x=(t,r,\vartheta,\varphi)=(x^0,x^1,x^2,x^3)$ with the matrices $\tilde\alpha^\mu$ given by \eqref{tildealpha}.
Here, the potential $V$ assumes values in the Hermitian matrices on $\CCC^4$ and includes the mass term $\beta m$, all connection coefficients that arise from covariant derivatives, and (if desired) an external electromagnetic field. The probability density in coordinates is then
\be
\rho_{\psi_t}(r,\vartheta,\varphi) =
\tilde{\jmath}^t(x) = \psi^\dagger (x)\, \tilde\alpha^t(x) \, \psi(x) = \frac{1}{\lambda(x)} \sum_{s=1}^4 |\psi_s(x)|^2\,.
\ee

\subsection{Probability Current in Configuration Space}

The wave function $\psi$ is a function on $\cup_{n=0}^\infty \cup_{\Sigma \in \foliation} \Sigma^n$, and its value at $(x_1,\ldots,x_n)$ lies in $\spin_{x_1} \otimes \cdots \otimes \spin_{x_n}$. In coordinates, we describe $\Sigma\in\foliation$ as $\Sigma = [0,\infty) \times \SSS^2$, so $\psi$ becomes a time-dependent function on $\cup_{n=0}^\infty \Sigma^n$, and its value at $(t,q)=(t,\vx_1,\ldots,\vx_n)$ lies in $(\CCC^4)^{\otimes n}$. The probability density in configuration space $\Sigma^n$ relative to coordinate volume (i.e., the $n$-particle analog of $\tilde{\jmath}^t$) is
\be
\rho_\psi(q) = \psi(q)^\dagger \psi(q) \prod_{j=1}^n \frac{1}{\lambda(\vx_j)}\,.
\ee
The current vector field $J$ in configuration space $\Sigma^n$ (relative to coordinate volume) has component $J^{ki}$ corresponding to $\vx^i_k$, the $i$-the coordinate of particle $k$, given by
\be
J^{ki}_\psi(q) = \psi(q)^\dagger \, \tilde\alpha^i_k(\vx_k) \, \psi(q) \, \prod_{j\neq k} \frac{1}{\lambda(\vx_j)}\,.
\ee 
The Bohmian motion is therefore given by \eqref{Bohm}, and the jump rate density by \eqref{ratedensity} with
\be 
J^\perp = J^{n,1}
\ee
for $\vx_n=(0,\omega)\in \sing \cap \Sigma$.

\subsection{Interior--Boundary Condition and Hamiltonian}

We impose on the wave function a condition at the singularity which we call an interior--boundary condition. Like usual boundary conditions, it is a condition on the values of the wave function at the boundary; unlike usual boundary conditions, it relates values on the boundary to values at interior points in other sectors of configuration space; see \cite{TT15a,TT15b,KS16,LSTT,co1} for recent discussions of interior--boundary conditions, and \cite{STT18,LN18} for such conditions for the Dirac equation.

Let $z$ be a point on $\sing  \cap \Sigma$ and $x_1,\ldots,x_{n-1} \in \Sigma \setminus \sing $. Let us first mention some common boundary conditions that are not appropriate for our purpose: First, for scalar wave functions, a Dirichlet boundary condition \cite{DD78,Wal80,ST04}
\begin{equation}
  \psi(x_1,\ldots,x_{n-1},z) =0
\end{equation}
is a common boundary condition; it leads to the (for our purposes uninteresting) behavior that no particle ever reaches the singularity. Second, for Dirac wave functions, common boundary conditions specify two of the four components of $\psi$ on the boundary, leaving the other two unspecified (e.g., \cite{FR15}).\footnote{More generally, for Dirac-type equations (i.e., first-order PDEs with matrix coefficients), the boundary condition specifies as many components of $\psi$ as the $\alpha$ matrix (the coefficient of the derivative) in the direction orthogonal to the boundary has positive eigenvalues \cite{STT18}.} The analog of the Dirichlet condition then sets half of the components to zero (as $\alpha^1$ has two positive and two negative eigenvalues), which again leads to zero current into the boundary \cite{STT18}. In contrast, the condition that we want should allow that there can be a probability current into the singularity but make sure that that current into the boundary of $\widehat\conf_n$ corresponds to the gain in probability in $\widehat\conf_{n-1}$.

To this end, we identify the singularity $\sing \cap \Sigma$ with $\SSS^2$ and choose a spinor field $N:\SSS^2\to \CCC^4$ with the property
\be
N^\dagger(\omega) \, \underset{0}{\alpha}^1 \, N(\omega) = 0
\ee
for all $\omega\in\SSS^2$. In coordinates, we may think of $\Sigma$ simply as $[0,\infty) \times \SSS^2$, and of $\psi^{(n)}$ as a function $\Sigma^n \to (\CCC^4)^{\otimes n}$; we write $q=(\vx_1,\ldots,\vx_n)$ for an $n$-particle configuration and $A_k$ for the $4\times 4$ matrix $A$ acting on the $k$-th spin index. 

The interior--boundary condition (IBC) reads
\be\label{ex3ibc}
(\underset{0}{\gamma}{}_{n+1}^{1}-iI_{n+1})  \, \psi^{(n+1)}\bigl(q, \omega\bigr)
=-\frac{i}{\hbar\sqrt{n+1}} \, \psi^{(n)}(q) \otimes (\underset{0}{\gamma}{}^{1}-iI) \, \underset{0}{\alpha}^1 N(\omega)
\ee
for every $n=0,1,2,\ldots$, $q\in \Sigma^n$, and $\omega\in\SSS^2$. Here, $\gamma^1$ with a 0 underneath means the standard $\gamma^1$ matrix, as opposed to the matrix representation (relative to some chosen basis) of the section $\gamma$ of $\CCC T\st \otimes \spin \otimes \spin^*$. See \cite{STT18} for a discussion of how to set up IBCs; condition \eqref{ex3ibc} is the analog of the IBC (8) in \cite{STT18}. 

The Hamiltonian acts according to the right-hand side of the Schr\"odinger equation
\begin{align}
i\hbar \frac{\partial \psi^{(n)}(q)}{\partial t} 
&= \sum_{k=1}^n \biggl( -i\hbar \sum_{i=1}^3 \tilde\alpha_k^t(\vx_k)^{-1}\tilde\alpha_k^i(\vx_k) \, \partial_{ki} \psi^{(n)}(q) + V_k(\vx_k) \, \psi^{(n)}(q) \biggr)\nonumber\\
&\quad + \sqrt{n+1} \int_{\SSS^2} \!\!\! d^2\omega\, N_{n+1}(\omega)^\dagger \, \psi^{(n+1)}\bigl(q,\omega \bigr)\label{ex3H}
\end{align}
at every interior point $q$. Here, $\int_{\SSS^2} d^2\omega$ means the same as $\int_0^{\pi} d\vartheta\, \sin^2\vartheta \int_0^{2\pi} d\varphi$, and the notation $N_{n+1}$ indicates that $N$ gets contracted with the last spin index of $\psi$.

\subsection{Conservation of Probability}

We carry out a calculation to check the unitarity of the time evolution and the equivariance of the Bohmian process $Q_t$. 

We write $H_0\psi$ for the first term on the right-hand side of \eqref{ex3H} and $H_I\psi$ for the second. The Schr\"odinger equation \eqref{ex3H} entails the continuity equation
\be\label{continuity3}
\frac{\partial\rho_{\psi_t}}{\partial t} (q) = -\sum_{k=1}^n \sum_{i=1}^3 \partial_{ki} J^{ki} 
+ \tfrac{2}{\hbar} \Im \,\psi^{(n)}(q)^\dagger \tilde{\alpha}^t_1(\vx_1) \cdots \tilde{\alpha}^t_n(\vx_n) \, (H_I \psi)^{(n)}(q) 
\ee
at $q=(\vx_1,\ldots,\vx_n)\in \conf_n^\circ$. 
In the absence of $H_I$ in \eqref{ex3H} (i.e., in the case $N(\omega)=0$), different sectors $\psi^{(n)}$ in Fock space decouple, leaving the free Dirac equation for $n$ particles, which is known to imply conservation of probability. So the first term in \eqref{continuity3} represents the transport of probability within the sector $\conf_n$, whereas the second term, which will be abbreviated as
\be
\frac{\partial\rho_{\psi_t}}{\partial t}\Big|_{H_I} 
\ee
in the following, is the contribution from $H_I\psi$ and couples $\psi^{(n)}$ to $\psi^{(n+1)}$. 
It represents a gain (or loss, if negative) of probability at $q$. On the other hand, there is a loss of probability in $\conf_{n+1}$ due to flux into the boundary of configuration space, and we will now show that this gain and loss compensate each other, leading to overall conservation of probability. 

At $\{q\cup\omega:\omega\in\SSS^2\}$, the probability current into the boundary is
\begin{align}
&-\int_{\SSS^2} \!\!\! d^2 \omega \, J^\perp(q\cup\omega)  \nonumber\\
&= -(n+1) \int_{\SSS^2} \!\!\! d^2\omega\, \psi^{(n+1)}(q,\omega)^\dagger \, \tilde\alpha^r_{n+1}(\omega) \, \psi^{(n+1)}(q,\omega) \, \prod_{j=1}^n \frac{1}{\lambda(\vx_j)} \nonumber\\
&= (n+1)\biggl( \prod_{j=1}^n \frac{1}{\lambda(\vx_j)}\biggr) \int_{\SSS^2} \!\!\! d^2\omega\biggl[ -\psi^{(n+1)}(q,\omega)^\dagger \, \underset{0}{\alpha}{}^1_{n+1} \, \psi^{(n+1)}(q,\omega)\biggr].
\label{currentintoS}
\end{align}
On the other hand, the gain term can be expressed as
\begin{align}
&\frac{\partial\rho_{\psi_t}}{\partial t}\Big|_{H_I} (q) \nonumber\\
&= \tfrac{2}{\hbar} \Im \,\psi^{(n)}(q)^\dagger \tilde{\alpha}^t_1(\vx_1) \cdots \tilde{\alpha}^t_n(\vx_n) \, \sqrt{n+1} \int_{\SSS^2} \!\!\! d^2\omega\, N_{n+1}(\omega)^\dagger \, \psi^{(n+1)}\bigl(q,\omega \bigr) \\
&= (n+1)\biggl( \prod_{j=1}^n \frac{1}{\lambda(\vx_j)}\biggr) \int_{\SSS^2} \!\!\! d^2\omega\, \tfrac{2}{\hbar\sqrt{n+1}} \Im \,\psi^{(n)}(q)^\dagger  N_{n+1}(\omega)^\dagger \, \psi^{(n+1)}\bigl(q,\omega \bigr).
\label{HIcurrent}
\end{align}
It suffices to show that the integrands in \eqref{currentintoS} and \eqref{HIcurrent} are equal. 
It was verified in \cite[Sec.~2.2]{STT18} that for any spinor $N\in\CCC^4$ with the property
\be
N^\dagger \, \underset{0}{\alpha}^1 N =0\,,
\ee
any complex number $\phi\in\CCC$, and any spinor $\chi\in\CCC^4$ with the property
\be\label{ex4ibc}
(\underset{0}{\gamma}^{1}-iI)  \, \chi
=-\tfrac{i}{\hbar\sqrt{n+1}} \, (\underset{0}{\gamma}{}^{1}-iI) \, \underset{0}{\alpha}^1 N \phi \,,
\ee
it follows that
\be
\tfrac{2}{\hbar\sqrt{n+1}} \Im \Bigl[ \phi^* N^\dagger \chi \Bigr] 
= - \chi^\dagger \, \underset{0}{\alpha}^1 \, \chi\,.
\ee
Setting, for fixed values of the spin indices $s_1,\ldots,s_n$ and fixed $q\in\Sigma^n$, $\phi = \psi^{(n)}_{s_1...s_n}(q)$ and $\chi_{s_{n+1}}= \psi^{(n+1)}_{s_1...s_{n+1}}(q,\omega)$, and then summing over $s_1,\ldots,s_n$, we obtain that the integrands in \eqref{currentintoS} and \eqref{HIcurrent} are equal. On the mathematically non-rigorous level, this corresponds to the conservation of probability and thus to a time evolution on Fock space that is unitary.

These considerations also yield the equivariance of $\rho_{\psi}$ for the Bohmian process $Q_t$: As pointed out already after \eqref{continuity2}, we need to show that \eqref{continuity2} and \eqref{continuity3} have the same form, i.e., that if $p=\rho_{\psi_t}$ at any one time $t$, then also $\partial_t p = \partial_t \rho_{\psi_t}$. Indeed, $p=\rho_{\psi_t}$ implies, by \eqref{ratedensity} and \eqref{Bohm2}, that 
\begin{align}
  \frac{\partial p}{\partial t}(q,t) 
  &= -\sum_{i=1}^{3n}\partial_i 
  J^i(q,t) -\int_{\SSS^2}d^2\omega\, J^\perp(q\cup \omega)^+ + \!
  \int_{\SSS^2} d^2\omega \,J^\perp(q,t)^- \\
  &= -\sum_{i=1}^{3n}\partial_i 
  J^i(q,t) -\int_{\SSS^2}d^2\omega\, J^\perp(q\cup \omega)\\
  &= -\sum_{i=1}^{3n}\partial_i J^i(q,t) +    
  \frac{\partial\rho_{\psi_t}}{\partial t}\Big|_{H_I} (q) =   
  \frac{\partial \rho_{\psi_t}}{\partial t}(q) \,,
\end{align}
as claimed.

\section{Comparison to Bell-Type Quantum Field Theories}
\label{sec:BTQFT}

Two ways of extending Bohmian mechanics to quantum field theory are known: either by postulating that a field configuration (rather than a particle configuration) is guided by a wave function (understood as a functional on the field configuration space) \cite{Bohm52,BH,Str07}, or by introducing particle creation and annihilation into Bohmian mechanics \cite{Bell86,crea1,crlet,crea2B,GT05b} (but see also \cite{CS07} for a third proposal). The second approach is called ``Bell-type quantum field theory,'' as the first model of this kind (on a lattice) was proposed by Bell \cite{Bell86}. In these theories, the motion of the configuration along deterministic trajectories is interrupted by stochastic jumps, usually corresponding to the creation or annihilation of particles. The jumps are governed by the following law prescribing the jump rate $\sigma^\psi$ (probability per time) in terms of the wave function $\psi$, which is usually from Fock space:
\begin{equation}\label{Bell}
  \sigma^\psi(q' \to q) = \tfrac{2}{\hbar} 
  \frac{[\mathrm{Im} \,\langle \psi | q \rangle \langle q | H_I | q' \rangle 
  \langle q' | \psi \rangle]^+}{\langle \psi | q' \rangle \langle q' | \psi \rangle} \,,
\end{equation}
where $H_I$ is the interaction Hamiltonian. The configuration $Q_t$ thus follows a Markov process in the configuration space $\conf$ of a variable number of particles, as defined in \eqref{confdef} above; see \cite{crea2B} for a detailed discussion of this process.

We expect that further analysis will show that our model for stochastic particle creation at a timelike singularity fits into the scheme of Bell-type quantum field theories when understood in the appropriate way. In other words, we expect that the jump rate \eqref{ratedensity} can be regarded as a limiting or generalized case of \eqref{Bell}.

\section{Requirements on the Time Foliation}
\label{sec:foliation}

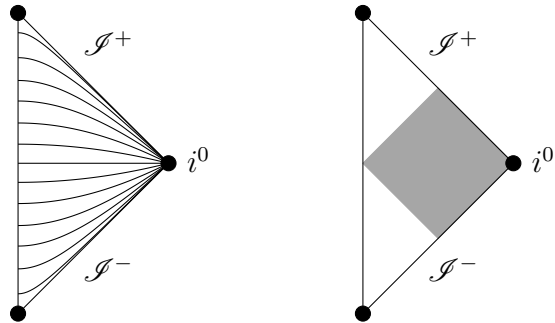
\begin{figure}[h]
\begin{center} 
  \begin{tikzpicture}
  \draw (0,0) -- (2,0);
  \draw [domain=0:2, samples=50] plot (\x,{2*(sqrt(0.02+\x*\x/4)-sqrt(1.02))});
  \draw [domain=0:2, samples=50] plot (\x,{2*(sqrt(0.13+\x*\x/4)-sqrt(1.13))});
  \draw [domain=0:2, samples=50] plot (\x,{2*(sqrt(0.4+\x*\x/4)-sqrt(1.4))});
  \draw [domain=0:2, samples=50] plot (\x,{2*(sqrt(1+\x*\x/4)-sqrt(2))});
  \draw [domain=0:2, samples=50] plot (\x,{2*(sqrt(3+\x*\x/4)-sqrt(4))});
  \draw [domain=0:2, samples=50] plot (\x,{2*(sqrt(15+\x*\x/4)-sqrt(16))});
  \draw [domain=0:2, samples=50] plot (\x,{-2*(sqrt(0.02+\x*\x/4)-sqrt(1.02))});
  \draw [domain=0:2, samples=50] plot (\x,{-2*(sqrt(0.13+\x*\x/4)-sqrt(1.13))});
  \draw [domain=0:2, samples=50] plot (\x,{-2*(sqrt(0.4+\x*\x/4)-sqrt(1.4))});
  \draw [domain=0:2, samples=50] plot (\x,{-2*(sqrt(1+\x*\x/4)-sqrt(2))});
  \draw [domain=0:2, samples=50] plot (\x,{-2*(sqrt(3+\x*\x/4)-sqrt(4))});
  \draw [domain=0:2, samples=50] plot (\x,{-2*(sqrt(15+\x*\x/4)-sqrt(16))});
  \draw (0,2) -- (0,-2);
  \draw (0,2) -- (2,0) -- (0,-2);
  \filldraw (0,-2) circle [radius=0.1];
  \filldraw (0,2) circle [radius=0.1];
  \filldraw (2,0) circle [radius=0.1];
  \node at (2.4,0.05) {$i^0$};
  %\node at (-0.8,0) {$r=0$};
  \node at (1.2,1.6) {$\mathscr{I}^+$};
  %\node at (1.8,1.1) {$r=\infty$};
  \node at (1.2,-1.4) {$\mathscr{I}^-$};
  %\node at (1.8,-1.1) {$r=\infty$};
  \end{tikzpicture}
\qquad\qquad
  \begin{tikzpicture}
  \draw[fill=gray!70!white,draw=gray!70!white] (0,0) -- (1,-1) -- (2,0) -- (1,1) -- cycle;
  \draw (0,2) -- (0,-2);
  \draw (0,2) -- (2,0) -- (0,-2);
  \filldraw (0,-2) circle [radius=0.1];
  \filldraw (0,2) circle [radius=0.1];
  \filldraw (2,0) circle [radius=0.1];
  \node at (2.4,0.05) {$i^0$};
  %\node at (-0.8,0) {$r=0$};
  \node at (1.2,1.6) {$\mathscr{I}^+$};
  %\node at (1.8,1.1) {$r=\infty$};
  \node at (1.2,-1.4) {$\mathscr{I}^-$};
  %\node at (1.8,-1.1) {$r=\infty$};
  \end{tikzpicture}
\end{center}
 \caption{LEFT: The foliation given by the $t$ coordinate function, shown in the Penrose diagram. RIGHT: The shaded region (comprising the points at spacelike separation from $(t=0,r=0)$) is where the $T$ function (timelike distance from the surface $\{t=0\}$) is finite, so that the level surfaces of $T$ foliate only the shaded region.}\label{fig:foliation}
\end{figure}

In the presence of timelike space-time singularities, difficulties arise about the choice of the time foliation that had not been encountered previously. While the foliation we used on the Reissner--Nordstr\"om space-time, given by the $t$ coordinate, worked well for our purposes, other foliations are problematical. Here is an example: Start with the $\{t=0\}$ hypersurface and propagate it according to \eqref{lawF} or $dn=0$, i.e., push every point of the hypersurface to the future at the same rate (so that the infinitesimal slice between two nearby hypersurfaces has constant proper thickness). The hypersurfaces we thus obtain are the level surfaces of the function $T$ that yields, for a space-time point $x$, the timelike distance of $x$ from the hypersurface $\{t=0\}$, i.e., the supremum of the lengths of the causal curves connecting $x$ to $\{t=0\}$.\footnote{To see this, note that the proper thickness of the infinitesimal slice between two nearby spacelike level surfaces of a function $f$, $\{f(x)=u\}$ and $\{f(x)=u+du\}$, is proportional to the proper length of the gradient of $f$, namely equal to $(\nabla^\mu f \nabla_\mu f)^{1/2} du$, because $\nabla^\mu f$ is orthogonal to the surface. The gradient of the $T$ function, where $T$ is differentiable, always has proper length 1 as a consequence of the definition of $T$.} At first sight, this might seem like a reasonable foliation, but in fact it does not cover all of $\st$ (or $\inte$), it only covers the shaded region in the right diagram of Figure~\ref{fig:foliation} (in contrast to the foliation defined by $t$, which covers all of $\st$ as visible in the left diagram of Figure~\ref{fig:foliation}). That is because the $T$ function is infinite in a large region of space-time, viz., outside the shaded region, which is the set of all points spacelike separated from $(t=0,r=0)$. Let me explain.

I begin by describing the outgoing radial null geodesics of the Reissner--Nordstr\"om geometry: They are the integral curves of the vector field $u^\mu = \lambda(r)^{-1/2}\partial_t + \lambda(r)^{1/2}\partial_r$, as this vector is null (easy to check), radial ($u^\vartheta=0=u^\varphi$), future-pointing ($u^t>0$), and outward-pointing ($u^r>0$). Thus, the outgoing radial null geodesics are explicitly given by
\be\label{orng}
t(r) = t_0 + \int_0^r \frac{dr'}{\lambda(r')}
\ee
with constant $\vartheta$ and $\varphi$. For small $r$, $t(r)-t_0 \approx r^3/3e^2$. Correspondingly, the \emph{incoming} radial null geodesics are given by
\be
t(r) = t_0 - \int_0^r \frac{dr'}{\lambda(r')}\,.
\ee
Note that at every $t_0\in\RRR$ and in every direction $(\vartheta,\varphi)$, one incoming radial null geodesic ends at $(t_0,r=0,\vartheta,\varphi)$ and one outgoing radial null geodesic begins.

Consider the curve \eqref{orng} for any $t_0>0$ and call it $C$. Any point $x$ on $C$ must have $T=\infty$: If it had a finite $T$ value, then choose $r_1$ so small that $\sqrt{\lambda(r_1)}>T(x)/t_0$ and $r_1 < r(x)$; that is possible because $\lambda(r)\to \infty$ as $r\to 0$. Let $x_1=(t(r_1),r_1,\vartheta(x),\varphi(x))$, which lies on $C$, too. Then the parallel to the $t$ axis (in coordinate space) through $x_1$ is a timelike curve connecting $x_1$ to the hypersurface $\{t=0\}$, and the proper time along this curve between $x_1$ and $\{t=0\}$ is $\sqrt{\lambda(r_1)} \, t(r_1)> \sqrt{\lambda(r_1)} t_0 > T(x)$. Since $T(x_1)$ is the supremum of such curve lengths, we have that $T(x_1)>T(x)$, so that $x$ must lie in the past of the spacelike hypersurface $\{T=T(x_1)\}$, in contradiction to the fact that $x$ lies on the future light cone of $x_1$.

Since $t_0>0$ was arbitrary, it follows that any point inside the future light cone of $(t=0,r=0)$ (i.e., in the future of the hypersurface defined by \eqref{orng} for $t_0=0$ and arbitrary $\vartheta,\varphi$) must have $T=\infty$. Likewise for the past light cone. And thus, $T$ foliates only the part of $\inte$ outside the light cone of $(t=0,r=0)$. Moreover, the leaves of $T$ cannot be extended to the singularity in such a way that their extensions remain mutually disjoint, as any $\{T=\mathrm{const.}\}$ hypersurface has $\{(t=0,r=0)\}\times \SSS^2$ as its boundary on the singularity. One could say that the leaves intersect on the singularity.

In contrast, the $t$ function provides a foliation of all of $\st$, including the singular boundary. That is a property we need of the time foliation.

\section{Conclusions}
\label{sec:conclusions}

We have set up a model of a quantum field theory in a curved space-time with a timelike singularity (the Reissner--Nordstr\"om geometry) that involves particle creation and annihilation at the singularity. The definition of the unitary time evolution involves an interior--boundary condition \eqref{ex3ibc} at the singular boundary of configuration space. We have also specified a stochastic process for the Bohmian configuration and verified that the appropriate $|\psi|^2$ distribution is equivariant. From the Bohmian perspective, it is particularly natural that particle creation and annihilation should occur at the singularity.

The rate of particle creation in the process for the Bohmian configuration is determined by some natural postulates. The interior--boundary condition is not unique; our choice is a particularly simple example.

The model suggests that naked timelike singularities are not as bad as they seem, although the model concerns not the stability of the Reissner--Nordstr\"om geometry but merely the behavior of test particles in a Reissner--Nordstr\"om background. That is because the model illustrates that, contrary to classical mechanics, the dynamics of particles need not be lawless in the presence of a naked timelike singularity.

\bigskip

\noindent\textit{Acknowledgments.} 
This research was supported by grant RFP1-06-27 from The Foundational Questions Institute (\url{fqxi.org}). I gratefully acknowledge helpful discussions with Shelly Goldstein, Michael Kiessling, Matthias Lienert, Frank Loose, Tobias Riedlinger, Julian Schmidt, Shadi Tahvildar-Zadeh, and Stefan Teufel.

\end{document}